\documentclass[A4,12pt]{article}
\usepackage{graphicx,url,latexsym,hyperref}
\usepackage[polish,english]{babel}
\usepackage[OT4]{polski}  

\newtheorem{theorem}{Theorem}

\newtheorem{observation}[theorem]{Observation}

\newtheorem{corollary}[theorem]{Corollary}

\newtheorem{definition}[theorem]{Definition}
\newtheorem{postulate}[theorem]{Postulate}
\newtheorem{proposition}[theorem]{Proposition}

\newcounter{examplectr}

\begin{document}

\begin{center}

{\Large \bfseries  On SAT information content,\\ 
its polynomial-time solvability \\
and fixed code algorithms}

\bigskip

{\large
\renewcommand{\thefootnote}{\fnsymbol{footnote}}

M.Drozdowski}

{\small
Institute of Computing Science,
Pozna\'{n} University of Technology,\\
Piotrowo 2, 60-965~Pozna\'{n}, Poland
}

\end{center}

\begin{abstract}
The amount of information in satisfiability problem (SAT) is considered.
SAT can be polynomial-time solvable when the solving 
algorithm holds an exponential amount of information.
It is also established that SAT Kolmogorov complexity is constant.
It is argued that the amount of information in SAT grows at least 
exponentially with the size of the input instance.
The amount of information in SAT
is compared with the amount of information in the fixed code algorithms
and generated over runtime.
\end{abstract}

\medskip

{\bfseries Keywords:}
computational complexity, 
information theory.

\section{Introduction}

A number of observations regarding the performance
of algorithms solving combinatorial problems and the amount of information 
they handle were made: 

\medskip

\noindent
$\bullet$ 
In \cite{FJ99} a connection between entropy of the Markov chains 
representing behavior of simulated annealing algorithms and the 
convergence of the expected objective function value has been made
for maximum 3-SAT problem.

\noindent
$\bullet$ 
In \cite{WM97} it is argued that there is a link between the fraction of 
problem instances achieving certain histogram of values and the entropy 
of the histogram.

\noindent
$\bullet$ 
In evolutionary optimization it is widely accepted rule of thumb that 
with growing population diversity and size, the chances of producing high 
quality solutions improve. 
Intuitively, such populations have more information.

\noindent
$\bullet$ 
Structural entropy of graphs representing linking between variables 
in SAT clauses correlates with practical solvability of SAT \cite{ZXZ21}.

\noindent
$\bullet$ 
There is a notion of a graph {\em hard-to-color} for a certain algorithm 
in graph node coloring \cite{K04,M97}.
A graph that is hard-to-color is colored by the considered algorithm
with more colors than the optimum.
There are examples of graphs hard-to-color for many deterministic algorithms.
Random sequential algorithm visits graph nodes in a random sequence
and assigns to a node the lowest feasible color.
For random sequential algorithm no hard-to-color graph exists, 
and hence, this algorithm cannot deterministically fail.
Ominously, this algorithm is connected to a source of randomness, that 
is, a source of unlimited amount of information.

\smallskip

In this paper we attempt to analyze the connection between computational
and information complexity of combinatorial problems.
The main result of this study is Theorem \ref{theo-SAT-expo-size}
and the following Corollary \ref{SAT-LinExpInfo} stating that the 
amount of information in SAT grows exponentially in the size of the 
input instance.
In the next section notions and assumptions are introduced.
Two reference points on the time-information space of SAT solvability 
are presented in sections \ref{sec-SAT-Tipoly-InfExp} and 
\ref{sec-SAT-TiExp-InfFix}.
In Section \ref{sec:Sat-Exp-Info} the main result on SAT information 
content is provided.
Section \ref{sec:in-conseq} is dedicated to the potential consequences of
the obtained SAT information amount estimate.

\section{Notions and Postulates}
\label{sec:define-postulate}

\begin{table}[t]
\caption{\label{tab-ntn} Summary of notations.}
\begin{center}
{\small
\begin{tabular}{|l|l|}
\hline

$\emptyset$ &
\parbox[t]{11cm}{symbol returned as a solution of a SAT "No"-instance}\\

$|A|$ &
\parbox[t]{11cm}{Size of algorithm $A$ in bits}\\

$D_\Pi$ &
\parbox[t]{11cm}{set of instances for problem $\Pi$}\\

$F$ &
\parbox[t]{11cm}{$F=\prod_{j=1}^mk_j$ conjunction of clauses $k_j$}\\

$F(I,\mathbf{\overline{x}})$ &
\parbox[t]{11cm}{value of $F$ for instance $I$ and bit assignment 
$\mathbf{\overline{x}}$}\\

$I$ &
\parbox[t]{11cm}{instance of a problem}\\

$|I|$ &
\parbox[t]{10cm}{instance size, i.e., length of the string encoding 
instance $I$ according to some reasonable encoding rule 
(e.g. numbers encoded at base greater or equal 2)}\\

$k_j$ &
\parbox[t]{11cm}{$j$th clause of SAT instance, for $j=1,\dots,m$}\\

$n$ &
\parbox[t]{10cm}{number of variables in the SAT problem}\\

$m$ &
\parbox[t]{10cm}{number of clauses in the SAT problem}\\

$\rho$ &
\parbox[t]{10cm}{upper bound on the bitrate of external 
information source, e.g., random bits source}\\

$S_\Pi(I)$&
\parbox[t]{10cm}{set of solutions for instance $I$ of search problem $\Pi$}\\

$x_i$ &
\parbox[t]{10cm}{$i$th variable in SAT problem, for $i=1,\dots,n$}\\

$\widetilde{x_i}$& 
\parbox[t]{10cm}{$i$th variable $x_i$ with or without negation}\\

$\mathbf{\overline{x}}$ &
\parbox[t]{11cm}{vector of $n$ binary values}\\

\hline
\end{tabular}
}
\end{center}
\end{table}

The notations used in this paper are summarized in Tab.\ref{tab-ntn}.
Search version of SAT problem is defined as follows:

\medskip

\noindent
{\sc SAT -- search version}

\noindent
{\sc Input:}
sums $k_j, j=1,\dots,m$, of binary variables, or their negations,
chosen over a set of $n$ binary variables $x_1,\dots,x_n$ .
The input data is SAT instance $I$.

\smallskip

\noindent
{\sc request:} 
Find the assignment of values 0/1 to binary variables $x_1,\dots,x_n$, 
i.e. vector $\mathbf{\overline{x}}$ of $n$ 0/1 values, such that the conjunction 
of the clauses $F(I,\mathbf{\overline{x}})=\prod_{j=1}^mk_j$ is 1. 
If such a vector does not exist then signal $\emptyset$.

\medskip

If the binary vector $\mathbf{\overline{x}}$ such that $F(I,\mathbf{\overline{x}})=1$ 
exists then we will be saying that $I$ is a "yes" instance. 
Otherwise $I$ is a "no" instance.
The input sums $k_j$ will be alternatively referred to as clauses.
If clauses $k_j$ comprise exactly three variables we will say that 
it is a 3-SAT problem instance.

\medskip

Let $\Sigma^+$ be a set of strings comprising instance encodings 
as well as solution encodings for some search problem, such as SAT,
using some reasonable encoding scheme $e$ over alphabet $\Sigma$. 
Search problems are string relations \cite{GJ79}:

\begin{definition}
A search problem $\Pi$ is a string relation 
\[R[\Pi,e]=
\left\{
(a,b): 
\parbox[c]{9cm}{$a\in\Sigma^+$ is the encoding of an instance
$I\in D_\Pi$ and $b\in\Sigma^+$ is the encoding of a solution 
$s\in S_\Pi(I)$ under coding scheme $e$
}
\right\},
\]
where $D_\Pi$ is a set of instances for problem $\Pi$ and
$S_\Pi(I)$ is a set of solutions for instance $I$ of $\Pi$.
\end{definition}

\smallskip
\noindent
Let $|I|$ denote instance $I$ size, i.e., length of the string encoding $I$ 
according to some reasonable rule.
For simplicity of the exposition we assume that $I$ is binary-encoded.
We will conventionally refer to bits as information amount units although
other units are also possible.

\begin{definition}
Fixed code algorithm is an algorithm which is encoded in
limited number of bits.
\end{definition}

\noindent
It is necessary to explain how size of an algorithm can be practically 
measured in bits.
Let us consider two models of algorithms: a Turing machine 
and a Random Access Machine (RAM).
Turing machine \cite{GJ79} is defined by 
set $\Gamma$ of tape symbols,
set $Q$ of states with distinguished halt states $Q_H$
and a transit function 
$\delta: (Q-Q_H)\times\Gamma\rightarrow Q\times\Gamma\times\{-1,+1\}$
determining for each pair of the current state in $Q-Q_H$
and read tape symbol from $\Gamma$, the next state in $Q$,
the symbol to be written on the tape and the direction of tape 
read-write head move.
The information content of all these objects is limited 
and an algorithm can be stored 
(e.g. in an array data structure, or a table of numbers)
in limited number of at most
$(\lceil\log|\Gamma|\rceil+\lceil\log|Q|\rceil+1)\times|\Gamma|\times|Q|$
bits.
Random Access Machine \cite{AHU74} has input and output tapes
and a program embodying the algorithm.
The program is a finite sequence of instructions from a limited instruction 
set and possibly some initial data (e.g. preset variables and constants 
defined by the programmer or the compiler).
Let $PI$ denote the number of program instructions from an instruction 
set of size $IS$.
The size of RAM program embodying some algorithm is upper-bounded
by $\lceil\log(IS)\rceil\times PI$ bits.
All data that a RAM program comprises at the outset of the computation
are also of limited size and can be counted in into the RAM code size.
RAM can be considered a simplified version of programs executed by 
the contemporary computers.
Hence, the programs executable on the contemporary CPUs that encode some 
abstract algorithms have limited information size.
The CPUs themselves comprise some algorithms which can be executed.
But likewise, CPUs need only a limited amount of information to be 
represented because the CPUs can be perceived as logical gates, 
connections between the gates and the microcode.
The number of program code bits and CPU representation bits is 
upper-bounding the information content of an algorithm.
Overall, the above objects representing algorithms can be described
in fixed number of bits.
Let $|A|$ denote the size of algorithm $A$ in bits.

Let us consider the relationship between fixed code algorithm, 
its data structures, deterministic and randomized algorithms.
An essential requirement for the further discussion is
that algorithm information contents size is upper-bounded by a constant.
A fixed code algorithm does not change its code size during the runtime.
The data-structures that the algorithm comprises at the outset of the 
computation may change its content during the runtime, but their sizes 
must remain fixed.
A fixed code algorithm can generate information as well as acquire
information from external sources.
The data obtained and created during runtime is not counted into 
the size of the fixed code algorithm size.
A fixed code algorithm can operate deterministically, but also
can use external source of information as, for example, 
a stream of random bits used in randomized algorithms.
Thus, there can be fixed code deterministic and fixed code randomized
algorithms.
These cases will be tackled in Section \ref{sec:in-conseq}.
Overall, the fixed code, deterministic and randomized algorithms are
different but non-disjoint types of objects.

\begin{postulate}[Information conservation postulate]
\label{theo:info-equivalence}
In order to solve a problem, an algorithm, an instance, algorithm states 
and other sources of information must be capable of representing 
at least the same amount of information as the amount of the information 
in the problem.
\end{postulate}

\begin{definition}
{\em Truly random bit sequence} (TRBS)
is a sequence of bits, that has no shorter representation.
\end{definition}

\noindent
In effect, a TRBS cannot be compressed, and the way to represent a TRBS of 
length $N$ bits, is to store it in its whole entirety on $N$ bits.

\begin{postulate}
\label{post-trbs-exist}
Truly random bit sequences exist.
\end{postulate}

\section{SAT Polynomial-time Solvability}
\label{sec-SAT-Tipoly-InfExp}

\begin{observation}
\label{obs-sat-Tipoly-Infoexp}
SAT can be solved in $O(|I|^2)$ time, {\em at least in principle}, 
by referring to precomputed solutions.
\end{observation}

{\bf Proof.}
Given instance $I$ of SAT, the statement of the above observation
can be expressed by the following pseudocode:

\medskip

\noindent
{\tt solution}$\leftarrow${\tt SolutionsTable[$I$];}

\medskip

\noindent
In the above peseudocode, SAT solution is retrieved from a table of 
precomputed solutions {\tt SolutionsTable} and the input instance $I$ 
is used as a position-index in {\tt SolutionsTable}.
Let us observe, that a simplifying assumption is often made that 
a table (an array) item can be referenced in constant time.
This simplification is not justified in the current case because
{\tt SolutionsTable} may be very large.
Therefore, and index data structure, in the sense of a database index,
is needed to prove that given $I$, its corresponding {\tt solution}
can be retrieved in $O(|I|^2)$ time.
To this end, the search for a precomputed solution of $I$ can be 
conducted in a binary tree with $2^{|I|}$ leaves and $2^{|I|}-1$ 
internal nodes using pointers (addresses) of length $|I|+1$ to arrive 
at the leaves.
An internal node holds two pointers to its successors.
A leaf holds an answer to a SAT instance 
(that is $\mathbf{\overline{x}}$ or $\emptyset$). 
The data-structure has size $O(2^{|I|}|I|)$ because it has
$2^{|I|+1}-1$ nodes each holding at most $2(|I|+1)$ bits
(which applies also to the leaves holding solutions).
The tree can be traversed top-down in $O(|I|)$ steps while reading 
instance $I$ bits.
Each read bit of $I$ determines whether the left, or the right, successor
of the current node is followed.
Retrieving the left or the right successor requires operation on a
$|I|+1$-bit-long addresses.
The total solution retrieval time is $O(|I|^2)$.

Thus, SAT can be solved in polynomial time, at least in principle,
provided that an algorithm for SAT has unlimited 
(precisely, exponential in $|I|$) amount of information about SAT.
$\hfill \Box$

\section{Kolmogorov Complexity of SAT}
\label{sec-SAT-TiExp-InfFix}

\begin{observation}
\label{obs-Kolmo-SAT}
SAT has constant Kolmogorov complexity.
\end{observation}

{\bf Proof.}
The minimum amount of information required to represent SAT 
as a string relation is at most $|E|+|V|$, where $E$ is an algorithm 
that enumerates all SAT instances $I$ and solutions according to some 
encoding scheme, while $V$ is an algorithm verifying if a given solution 
for $I$ is correct.
The proof that fixed code $E$ and $V$ exist is very technical because 
it refers to serialized representation of numbers for which addition 
can be executed by fixed size code without a need for a circuitry
which complexity grows with the values of the numbers.
In the following we outline key elements of $E$ and $V$ operation.

For simplicity of the exposition algorithms $E$ (enumerator) and 
$V$ (verifier) will be represented as two Turing machines, 
while $V$ has read access to the tapes of $E$.
Let us consider certain number $n=1,\dots$ of binary variables.
Given $n$, a SAT instance may be encoded as a sequence of values:
$(n, m,k_1,\dots,k_m)$. 
Since each variable can be present or missing in a clause, and if present 
in a clause, the variable can be used with or without negation,
the number of possible clauses is at most $2^{2n}$ and $m$ can be encoded 
in at most ${2n}$ bits.
Each clause can be encoded as a sequence of $n$ bit pairs representing 
at position $i=1,\dots,n$: 
$00_i$ or $01_i$ -- variable $x_i$ is absent in the current clause,
$10_i$ -- variable $x_i$ is present in the current clause as $x_i$ 
(without negation),
$11_i$ -- variable $x_i$ is present in the current clause as 
$\overline{x_i}$ (with negation).
Thus, each clause can be encoded in $2n$ bits.
The clauses of the instance can be encoded in 
$2nm\leq 2n\times 2^{2n}$ bits.
The whole SAT instance can be encoded as a binary number of length
$\lfloor\log n\rfloor+1+{2 n}+2nm$ bits.
Similarly, it is possible to enumerate all $2^n$ potential solutions 
$\mathbf{\overline{x}}$ 
of a SAT instance with $n$ variables on an $n$-bit-long binary string. 

All possible pairs $(I,\mathbf{\overline{x}})$ can be enumerated by 
a constant information size Turing machine (i.e. a fixed code algorithm)
$E$ with four tapes:
tape 1 holding value $n$ (this corresponds to the outermost enumeration loop),
tape 2 holding $m$ (2nd outermost loop), 
tape 3 holding $(k_1,\dots,k_m)$ (3rd outermost loop), and
tape 4 holding $\mathbf{\overline{x}}$ (the innermost loop).
The tapes extend to infinity in both directions.
Ends of the information on the tapes in $E$ and $V$ can be sensed by 
reading a special blank symbol "$b$".
The found pairs $(I,\overline{x_i})$ or $(I,\emptyset)$ of SAT as a string 
relation, are stored on the 5th tape as quadruplets:
$n, m, (k_1,\dots,k_m),\mathbf{\overline{x}}$, 
i.e. contents of tapes 1, 2, 3, 4, is copied, separated by and ending with 
symbol $b$.  
At the outset of the computation tapes 1, 2, 3, 4 comprise only blank 
symbols $b$, tape 5 comprises symbols $\dots,b,s,b,\dots$ to mark 
that no pair from the string relation has been found.
The Turing machine $E$ is working on the following principles:

\noindent
(1) 
It is adding 1 to a binary-encoded $\mathbf{\overline{x}}$ on tape 4.
Such addition can be implemented with a 3-state, 9-arc transit function.
If the end of tape 4 is reached, that is 1 is successfully added, 
the second Turing machine $V$ is called 
(it is presented in the following) with the tapes 3 and 4 of $E$ as an input.
If $V$ returns that $\mathbf{\overline{x}}$ on tape 4 satisfies 
the formula encoded on tape 3 then the content of tapes 1,2,3,4 
is copied to tape 5 as described above. 
$E$ proceeds to the beginning of (1).

\noindent
(2) 
If the 4th tape overflows, that is all $2^n$ possible values of 
$\mathbf{\overline{x}}$ are enumerated, then machine $E$ 
reads the symbol on tape 5.
If it is $b$ then this symbol is replaced with
$s$ to mark the end of the block of solutions for the instance 
currently encoded on tapes 1, 2, 3.
If symbol $s$ is read from tape 5, then no solution has been found for
the current instance, and contents of tapes 1,2,3 is copied to tape 5,
after which symbols $\emptyset, s$ are appended.
Next, $E$ adds 1 to number 
$(k_1,\dots,k_m)$ on tape 3 and returns to the enumeration 
of $\mathbf{\overline{x}}$ on tape 4 (step 1).

\noindent
(3)
If tape 3 overflows (i.e. exceeds $2nm$), then $2n$ zeros are 
appended on tape 3 to the string encoding $(k_1,\dots,k_m)$,
this operation can be facilitated by referring to the length of the string 
encoding $m$ on tape 2, as it also has length of $2n$ bits.
Furthermore, $m$ is increased by one.
If $m$ on tape 2 does not overflow, $E$ returns to (1).

\noindent
(4)
If $m$ overflows (i.e. exceeds $2^{2n}$), then bits 10 are appended to 
the binary encoding of $m$ and $n$ is increased on tape 1.

\noindent
Thus, a fixed code algorithm $E$ enumerating all input instances and all 
solutions for arbitrary $n$ exists.

It remains to show that algorithm $V$ checking if the binary-encoded 
$\mathbf{\overline{x}}$ on tape 4 satisfies formula $(k_1,\dots,k_m)$ on 
tape 3 can be implemented in a in transit function 
with limited number of states and arcs.
$V$ has two read heads for tapes 4 and 3 of $E$, the tapes are read from 
left to right, and while reading literals in clauses $(k_1,\dots,k_m)$ 
on tape 3 it moves the read head on tape 4 accordingly.
In detail, $V$ operates on the following principles:

\noindent
state 0: 
start with the read heads at the beginnings of tapes 3 and 4 
(the leftmost positions).

\noindent
state 1: \\
$\bullet$ 
If head 4 reads $b$ (end of tape 4, values of all binary variables were
verified, but none satisfied the current clause $k_j$ on tape 3),
move both heads to the beginnings o the tapes,
return to $E$ to the state accepting an answer that 
$\mathbf{\overline{x}}$ on tape 4 does not satisfy formula 
$(k_1,\dots,k_m)$ on tape 3 (i.e. return to the beginning of $E$ point (1)).
Otherwise head 4 reads 0/1, then do the following:

\noindent
$\bullet$ 
If head 3 reads 0 (the current variable $x_i$ is not present in the 
current clause $k_j$), then tape 3 is moved by two and tape 4 by one 
position to the right, next jump to state 1.

\noindent
$\bullet$ 
If head 3 reads 1 ($x_i$ is present in $k_j$), 
then move tape 3 one position to the right, and proceed to state 2.

\noindent
$\bullet$ 
If head 3 reads $b$ ($b$--blank, end of the tape 3 is reached,
all clauses satisfied)
move heads 3 and 4 to the beginning of the tapes
and return to $E$ to the state accepting "yes" answers.

\noindent
state 2: \\
$\bullet$ 
If 
head 3 reads 0 and head 4 reads 1, or 
head 3 reads 1 and head 4 reads 0 ($k_j$ is satisfied by $x_i$), 
then iteratively move head 4 by one position to the right and head 3 two 
positions to the right at an iteration, until reaching end of tape 4
(head 3 is at the start of clause $k+1$ or moved beyond the end of tape 3).
Move head 4 to the beginning of the tape.
Proceed to state 1.

\noindent
$\bullet$ 
Otherwise, ($k_j$ is not satisfied by $x_i$)
move both heads one position to the right and proceed to state 1.

\smallskip

\noindent
Hence, SAT as a string relation can be reconstructed by enumerating 
all input instances of increasing sizes $n$ using algorithm $E$ and 
choosing the correct answer by the use of algorithm $V$.
$\hfill \Box$

\bigskip

It is an interesting coincidence that SAT is polynomial-time solvable 
if exponential amount of information is held by an algorithm 
(Observation \ref{obs-sat-Tipoly-Infoexp})
and a fixed-size information object is sufficient to recreate
SAT in exponential time (Observation \ref{obs-Kolmo-SAT}).
Note that information in one form of SAT representation, that is
concise $E$ and $V$, is transformed to different information
in alternative SAT representation (string relation).

\section{Amount of Information in SAT}
\label{sec:Sat-Exp-Info}

In this section we introduce the idea of quantifying information content 
of search problems as string relations.
A string relation $R[\Pi,e]$ is a mapping from strings $a$ representing 
instances to strings $b$ representing solutions.
The information content of problem $\Pi$ can be measured
in the terms of the amount of information carried by this mapping.

SAT-search is also an example of a search problem.
Thus, also SAT can be thought of as a mapping from strings $a$ representing 
instances to strings $b$ representing solutions and this mapping
can be measured in the terms of the amount of carried information.
Each string $a$ is either a "yes" instance, or a "no" instance. 
In the former case an $n$-bit solution $\mathbf{\overline{x}}$ must be provided by 
the mapping.
In the latter case symbol $\emptyset$ must be provided.
We will assume conventionally that if the $a$ string, 
according to encoding scheme $e$, 
is not encoding any SAT instance, then such a case can 
be represented in the same way as a "no" instance with answer $\emptyset$.
In order to encode each pair $(a,b)$ of the relation representing SAT
it is necessary to have an equivalent of a graph arc from string $a$
to its solution $b$.
Such an arc requires $|I|+n$ bits of information which is at 
least $\Omega(|I|)$ bits.
There are $2^{|I|}$ strings of some size $|I|$.
Since it is necessary to at least distinguish whether $b$ strings
represent $\emptyset$ or $\mathbf{\overline{x}}$,
at least $\Omega(2^{|I|})$ bits of information seem necessary 
to encode SAT as string relation $R[SAT,e]$.
However, it is still possible that SAT can be encoded in fewer than 
$\Omega(2^{|I|})$ bits.
Thus, some more compact, or compressed, representation of SAT
may exist.

\begin{theorem}
\label{theo-SAT-expo-size}
The amount of information in SAT grows at least exponentially
with instance size.
\end{theorem}

{\bf Proof.}
Assume there are $n$ variables and $4n$ clauses in 3-SAT.
Let there be 4 clauses 
$
k_{i1}=x_a+x_b+\widetilde{x_i},
k_{i2}=\overline{x_a}+x_b+\widetilde{x_i},
k_{i3}=x_a+\overline{x_b}+\widetilde{x_i},
k_{i4}=\overline{x_a}+\overline{x_b}+\widetilde{x_i}
$
for each $i=1,\dots,n$.
$\widetilde{x_i}$ denotes that variable $x_i$ will be set with 
or without negation.
No valuing of $x_a,x_b$ makes the four clauses simultaneously equal 1.
The four clauses may simultaneously become equal 1 only if $\widetilde{x_i}=1$.
Satisfying formula $F=k_{11}k_{12}k_{13}k_{14}\dots k_{n4}$ depends on 
valuing of variables $\widetilde{x_i}$ for $i=1,\dots,n$.
Depending on whether binary variable $i$ is negated or not 
(i.e. written either $x_i$ or $\overline{x_i}$ consistently in 
$k_{i1},k_{i2},k_{i3},k_{i4}$)
there can be $2^n$ different 
ways of constructing formula $F$, thus leading to $2^n$ different 
"yes" instances with $2^n$ different solutions.
Variables $x_a, x_b$ are chosen such that $a\neq b$ and $a,b\neq i$.
Since there are $(n-1)(n-2)/2$ possible pairs $a,b$ for each $i$, 
it is possible to generate pairs $a,b$ satisfying the above conditions 
for $n\geq 3$.

We are now going to calculate the number of different "yes" 
instances as a function of instance size $|I|$
because instance size $|I|$, not $n$, is used in the complexity
assessment.
Suppose the uniform cost criterion \cite{{AHU74}} is assumed, 
then each number has value limited from above by constant $K$.
The length of the encoding of the instance data is 
$|I|=4n\times 3\log K+\log K=12n\log K+\log K$ because it
is necessary to record the indexes of variables in $\log K$ bits, 
each binary variable induces 4 clauses of length $3\log K$.
Negation of a variable, or lack thereof, is encoded on one bit
within $\log K$.
Consequently, the number of possible unique solutions is 
$
2^{n}=2^{(|I|-\log K)/(12\log K)}=
2^{|I|/(12\log K)}2^{-1/12}$,
which is $\Omega(2^{d_1|I|})$, where $d_1=1/(12\log K)>0$.

Assume logarithmic cost criterion \cite{{AHU74}}, then the number of bits 
necessary to record $n$ is $\lfloor\log n\rfloor+1$,
and $\lfloor\log n\rfloor+2$ bits are needed to encode the index 
of a variable and its negation, or lack thereof.
Length of the encoding string is 
$|I|=12n(\lfloor\log n\rfloor+2)+\lfloor\log n\rfloor+1\leq
15n\log n=dn\ln n$, for $n>2^{24}$ and $d=15/\ln 2\approx 21.6404$.
An inverse function of $(cx\ln x)$, for some constant $c>0$, is
$\frac{x}{c}/W(\frac{x}{c})$, where $W$ is Lambert $W$-function
\cite{WA-Lambert}.
Lambert $W$ function for big $x$ can be approximated by
$W(x)=\ln x-\ln \ln x+O(1)$.
Given instance size $|I|$, we have
$n\geq \frac{|I|}{d}/W(\frac{|I|}{d})\approx 
\frac{|I|}{d}/(\ln\frac{|I|}{d}-\ln\ln\frac{|I|}{d}+O(1))\geq
\frac{|I|}{d}/(2\ln\frac{|I|}{d})\geq
\frac{|I|}{d}/(2\ln |I|-2\ln d)\geq
|I|/(2d\ln |I|),
$
for sufficiently large $|I|$.
Note that $|I|, dn\ln n, \frac{x}{c}/W(\frac{x}{c})$ are increasing in $n,x$.
Thus, by approximating $|I|$ from above we get a lower bound of $n$ after 
calculating an inverse of the upper bound of $|I|$.
The number of possible unique solutions is 
$2^{n}\geq 2^{|I|/(d_2\ln |I|)}$ where $d_2=2d$.
Observe that $2^{|I|/(d_2\ln |I|)}$ exceeds any polynomial function of 
$|I|$ for sufficiently large $|I|$, because for any polynomial function 
$O(|I|^k)$, $\ln(|I|^k)<|I|/(d_2\ln |I|)$ with $|I|$ tending to infinity.

Consider a truly random bit sequence (TRBS) of length $2^n$.
Assume that $j\in\{1,\dots,2^n\}$ is one of the instances of 3-SAT 
constructed in the above way.
Let $j[i]$ for $i=1,\dots,n$ be the $i$-th bit of $j$ binary encoding.
If bit $j$ of the TRBS is equal to 1, we set variables
$\widetilde{x_{i}}$, for $i=1,\dots,n$, such that $\widetilde{x_{i}}=j[i]$ 
satisfy clauses $k_{1i},\dots,k_{4i}$.
For example, if $j[i]=1$ and the $j$-th bit of the TRBS is 1, 
then $\widetilde{x_{i}}$ is written as $x_{i}$.
If $j[i]=0$ and the $j$-th bit of the TRBS is 1, 
then $\widetilde{x_{i}}$ is written as $\overline{x_{i}}$.
Thus, if bit $j$ of the TRBS is equal to 1 then a "yes" instance is constructed.
Conversely, if the TRBS bit $j=0$ then at least one variable $x_i$ in the 
corresponding clauses $k_{1i},\dots,k_{4i}$ is set inconsistently, i.e.,
some $\widetilde{x_i}$ appears in $k_{1i},\dots,k_{4i}$ both with negation
and without.
Hence, if bit $j$ of the TRBS is equal to 0, the $j$-th instance
constructed in the above way becomes a "no" instance.
Note that in this way the TRBS of length $2^n$ was encoded 
in 3-SAT search problem.
The amount of information in 3-SAT grows at least in the order of 
$\Omega(2^{d_1|I|})$ for uniform 
(or $\Omega(2^{|I|/(d_2\ln |I|)})$ for logarithmic)
cost criterion.
$\hfill \Box$

\medskip

Let us observe that there can be $2^{2^n}$ different TRBSes
of length $2^n$ used in the proof of Theorem \ref{theo-SAT-expo-size}.
However, it does not mean that $2^{2^n}$ SAT examples 
(i.e. batches of $2^n$ SAT instances constructed as described 
in Theorem \ref{theo-SAT-expo-size}) can be delivered.
Note that in the construction of Theorem \ref{theo-SAT-expo-size}
it is important whether instance $j\in\{1,\dots,2^n\}$ is a 
"yes", or a "no" instance.
After presenting the $2^n$-bit-long TRBS, only when instance 
$j\in\{1,\dots,2^n\}$ is presented in the alternative form to the TRBS
a new piece of information emerges.
That is, if instance $j$ was a "yes" instance
in the construction of Theorem \ref{theo-SAT-expo-size},
then a new information on $j$ is that it can be also made a "no" instance.
And vice versa, new information emerges when some "no" instance $j$
in the TRBS of Theorem \ref{theo-SAT-expo-size} is later presented 
as a "yes" instance.
Thus, information on the permutation in which alternative versions
of instances $j\in\{1,\dots,2^n\}$ are revealed can be added to 
the $2^n$ bits of the Theorem \ref{theo-SAT-expo-size} TRBS.
Since there are $2^n!$ permutations of instances $j$ alternative forms,
the additional amount of information in their permutation is 
$\lfloor\log 2^n!\rfloor+1$.
This results in SAT information estimation
$\Omega(n2^n)$ which is $\Omega(|I|2^{d_1|I|})$ for uniform 
(or $\Omega(|I|/(d_2\ln |I|)2^{|I|/(d_2\ln |I|)})$ for logarithmic)
cost criterion and the construction used in 
Theorem \ref{theo-SAT-expo-size}.
Conversely, the amount of information necessary to 
record directly all $2^n$ solutions to the instances of length $|I|$
with $n$ binary variables is $O(|I|2^n)$. 
Since $n\leq |I|$ (for both types of cost criteria) we have a corollary:

\begin{corollary}
\label{SAT-LinExpInfo}
Information content of SAT is $\Theta(|I|2^{|I|})$.
\end{corollary}

\section{On the Consequences}
\label{sec:in-conseq}

\subsection{Polynomial-time Solvability and Information Size}
\label{sec:PolyTiSolv-vs-Info}

Let us estimate the amount of information that can be created
by a fixed code algorithm running in time $T$.

\begin{proposition}
\label{prop-own-info}
A fixed code deterministic algorithm can produce $\Omega(T\log T)$ 
bits of information in $T$ units of time.
\end{proposition}

{\bf Proof.}
The information represented in the evolution of the algorithm state
can be estimated in at least three ways, all leading to the same
lower bound.
The number of algorithm different states is $T$ because 
the algorithm stops and hence it does not loop.
The number of different sequences in which the states of the algorithm
can be visited can be bounded from above by $T!$.
The number of bits necessary to distinguish the sequences is
$\lfloor\log(T!)\rfloor+1$ which is $\Theta(T\log T)$.
An attempt of distinguishing the states of the algorithm
by counting them, or assigning time stamps, gives only 
$T\log T$ bits of information.
It is also possible to assess the amount of information in the evolution 
of the algorithm state by use of mutual information 
between the progressing time and algorithm state.
Assume that time variable ${\cal T}$ progresses deterministically 
in steps $1,\dots,T$, then the probability that step $t$ is achieved 
by time $t$ is $p(t)=1$.
The algorithm state variable ${\cal S}$ progress through states 
$s_1,\dots,s_T$.
The deterministic algorithm is in certain state $s_t$ at time $t$
with probability $p(s_t,t)=1$ and $\forall s_i\neq s_t, p(s_i,t)=0$.
The probability that algorithm is in state $s_t$ disregarding time
is $1/T$.
Mutual information from time variable ${\cal T}$ to algorithm state 
${\cal S}$ is 
$
I({\cal S},{\cal T})=
\sum_{i=1}^T\sum_{s_i\in{\cal S}}p(s_i,i)\log\frac{p(s_i,i)}{p(s_i)p(i)}=
\sum_{i=1}^Tp(s_t,t)\log\frac{p(s_t,t)}{p(s_t)p(t)}=
\sum_{i=1}^T1\times\log\frac{1}{1/T\times 1}=
T\log T.
$
$\hfill \Box$

\bigskip

According to the information conservation postulate 
\ref{theo:info-equivalence},
the input instance, the algorithm,
the information derived from changing algorithm states 
and obtained from external sources over the runtime must represent 
equivalent amount of information as the mapping 
from the instances to the solutions in the considered problem.
An algorithm solves a problem if it provides an answer 
{\em for each} input instance \cite{GJ79}.
However, at the outset of computation only one instance is given, 
and the amount of information in a fixed code algorithm and in the input 
instance is $|I|+|A|$ bits.
We argue that the amount of information derived from changing algorithm 
states or from external sources can be insufficient for some
algorithm classes.
Let $\rho$ denote an upper limit on the amount of information 
that can be transferred in a unit of time ($\rho$ is a bitrate limit).
The assumption on upper-bounding of $\rho$ is practical because it is not 
possible to transfer arbitrary amounts of information in limited time.
The bitrate limit applies both to the transfers of the algorithm 
internally created information and to the transfers of external 
information, e.g. a stream of random bits, which can be acquired 
by an algorithm.

\begin{proposition}
\label{theo:poly-incapable-with-xtra}
Fixed code algorithm is not capable of representing SAT
in polynomial time even with external source of information
of constant bitrate.
\end{proposition}

{\bf Proof.}
$|I|+|A|$ is the instance and the algorithm information size.
By proposition \ref{prop-own-info} the amount of information 
created by a fixed code algorithm run in time $T=q(|I|)$, 
where $q$ is a polynomial, is $\Omega(q(|I|)\log(q(|I|)))$.
The amount of acquired external information is $O(\rho q(|I|))$.
Also no more than $O(\rho q(|I|))$ information can be transferred
by the algorithm internally, which applies also to the information 
created by the algorithm itself. 
Thus, for sufficiently large $|I|$ the amount of available information 
an algorithm is able to use is limited by $O(\rho q(|I|))$
rather than by the size $\Omega(q(|I|)\log(q(|I|)))$
which could potentially be created.
The information amount accessible for a fixed code algorithm 
for sufficiently large $|I|$ is $|I|+|A|+O(\rho q(|I|))$.
Overall, for sufficiently large $|I|$ it is less information than 
$\Omega(2^{|I|})$ bits comprised in SAT by 
Theorem \ref{theo-SAT-expo-size}.
$\hfill \Box$

\bigskip

Let us return to Kolmogorov complexity of SAT and SAT size 
as a string relation.
On the one hand, Kolmogorov complexity of SAT is $|E|+|V|$, 
by Observation \ref{obs-Kolmo-SAT}. 
On the other hand, by Theorem \ref{theo-SAT-expo-size} 
SAT has $\Omega(2^{|I|})$ (uniform criterion) or 
$\Omega(2^{|I|/(\ln |I|)})$ (logarithmic criterion) 
incompressible bits.
It can be speculated that the discrepancy between these two numbers can be 
attributed to qualitative difference of the two forms of SAT representation
(two types of information).
Transforming from the first form to the second 
requires $\Omega(2^{|I|})$ information derived over runtime.
Informally, SAT has exponential compression efficiency 
with respect to $|I|$.
Since by Cook's theorem SAT is a foundation of all {\bf NP}-complete 
problems, the above observations can be extended to all 
{\bf NP}-complete problems.

\subsection{On Polynomial Problems Information Content}
\label{sec:PolyTiSolv-ContentInfo}

A complement to the considerations in Section \ref{sec:PolyTiSolv-vs-Info}
is to show that pol\-y\-no\-mial-time-solvable problems have 
polynomial amount of information when represented as string relations.
Unfortunately, a uniform and universal approach to achieve this 
is unknown.

As an attempt to study a string relation representing a polynomial problem 
and its information content, consider sorting $n$ integers $l_i$.
Input instances $a$ can be any of the $n!$ permutations of the numbers.
The output string $b$ is the sequence satisfying
$l_1\leq l_2,\dots\leq l_n$.
Note that the actual values of $l_i$s are immaterial.
The actual input permutation of the numbers is also meaningless.
Important is the fact that relations $l_i\leq l_j$ can be established.
The amount of information required to establish this fact is
$|l_i|+|l_j|+|cmp|$, where $|l_i|$ is binary-encoded string length of $l_i$,
$|cmp|$ is the size of a fixed code comparator algorithm.
Such a comparator can be implemented as a 2-tape Turing machine
with 3 states (including two finals states: $l_i\leq l_j$, $l_i>l_j$), or
with 5 states if replacing positions of the numbers on the tapes is needed.
Hence, the input information size is $|I|$ on the side of input strings $a$
of the string relation, independently of the numbers permutation.
The amount of information on the side of the output strings $b$ satisfying
$l_1\leq l_2,\dots\leq l_n$ is $O(n\log n)$ which can be upper-bounded by 
$O(|I|)$.
In order to confirm that sorting numbers, transforming any input 
permutation $a$ to the required sequence of numbers $b$, carries 
polynomial amount of information, consider, e.g., bubble-sorting network 
which has $O(n^2)$ comparators and $O(\log n)$ bits of information are 
needed to connect inputs of each comparator to the outputs of its 
predecessor in the sorting network.
The sorting network has $O(n^2\log n)$ information which can be
upper-bounded by $O(|I|^2)$.
The overall information in sorting grows in $O(|I|^2)$
with the instance size.

The above reasoning can be extended to all combinatorial problems 
which solutions are obtained by sorting input elements
according to some rule, that is, to the problems solvable 
by greedy algorithms.

\subsection{On Non-fixed Code Algorithms}

In this study a concept of fixed code algorithms was used.
A question naturally emerges, what the non-fixed code algorithms can be.
Classic randomized metaheuristics, although they use external sources of
randomness, remain within the realm of fixed code algorithms.
Machine learning methods seem obvious candidates for the non-fixed code 
algorithms if the training and the inference stages together are 
considered one algorithm.
The inference stage, when used stand-alone, also remains in the 
realm of fixed code algorithms.
Let us note that by Theorem \ref{theo-SAT-expo-size}, 
combinatorial problems like SAT, are ultimately not learnable.
This poses a question of the scalability of machine learning methods
in solving hard combinatorial problems, possibility of machine learning
auto-tuning and taking humans out of the training-inference loop.

\section*{Acknowledgments}

I thank Joanna Berli\'nska, Iwo B\l \k{a}dek, 
Piotr Formanowicz,
Ma\l gorzta Sterna,
Tomasz \.Zok,
Krzysztof Zwierzy\'nski
for discussing with me in the earlier
stages of this consideration \cite{D16, D23}.
To Patrick de Causmacker, Seffi Naor, Natasha Shakhlevich,
Stanis\l aw Gawiejnowicz 
for inspiring questions.

\end{document}